\documentclass[twocolumn,showpacs,prb,amsmath,amssymb]{revtex4}
\usepackage{graphicx}
\usepackage{bm}
\begin{document}

\title{The Exchange Gate in Solid State Spin Quantum Computation: The
  Applicability of the Heisenberg Model}
\author{V.W. Scarola and S. Das Sarma}
\affiliation{Condensed Matter Theory Center, 
Department of Physics, University of Maryland,
College Park, MD 20742-4111}

\begin{abstract}

Solid state quantum computing proposals rely on adiabatic
operations of the exchange gate among localized spins in
nanostructures.  We study corrections to the 
Heisenberg interaction between lateral 
semiconductor quantum dots in an external magnetic field.  Using exact
diagonalization we obtain the regime of validity of the adiabatic
approximation.  We also find qualitative corrections to the
Heisenberg model at high magnetic fields and in looped arrays of
spins.  Looped geometries of localized spins generate flux dependent,
multi-spin terms which go beyond the basic Heisenberg model.

\end{abstract}
\pacs{03.67.Pp, 03.67.Lx}
\maketitle

\section{Introduction}
Scalable quantum computation proposals, particularly in solid state
architectures compatible with existing microelectronic technology, are 
of great potential importance.  The exchange gate \cite{Loss}, based on the
Heisenberg interaction between localized spins is a key concept
underlying several proposed quantum computer architectures in
semiconductor nanostructures, where the spin of a localized electron
serves as the single qubit.\cite{Loss,Hu,Kane,Vrijen}  In most 
spin-based quantum computer
proposals single-qubit operations involve rotations of individual 
spins by external magnetic field pulses
(essentially a modified electron spin resonance technique
implemented on individual spins) or by some other techniques ({\em e.g.} 
local g-factor manipulation by external electric field pulses).  The
two qubit operation in solid state, spin architectures is projected to
be achieved by the exchange gate operation.  The ability of a tunable 
exchange gate (which enables the exchange
coupling to change from zero to a finite value within in a
``reasonable'' gating time) in carrying out the ``SWAP'' operation
leads to the implementation of the universal 2-qubit CNOT gate which, 
along with single qubit gates, provide a universal set of quantum
gates.  The perceived advantages of solid state quantum computation
are its scalability (because of the existing semiconductor
microelectronics infrastructure), the low decoherence rate for spin
states (as compared for example, with charge states in
semiconductors), and the feasibility of precise control over the local
inter-electron Heisenberg coupling through the exchange gate.
Among the disadvantages is the inevitable presence
of some spin decoherence due to the background solid state
environment \cite{DeSousa} and difficulties in the
measurement of single electron spin states.  However, recent experimental 
advances \cite{detection} have demonstrated single spin measurements
of localized electrons embedded in semiconductors.

In this work we provide a detailed and quantitatively accurate
theoretical study of two complementary aspects of solid state spin
quantum computer architectures.  First, we develop a theory for
obtaining an accurate map of the low lying Hilbert space of the
quantum dot based spin quantum computer architecture (the
Loss-DiVincenzo\cite{Loss} architecture in coupled GaAs quantum dots) in the
presence of an external magnetic field.  Our calculation of the 
Hilbert space structure of the coupled quantum dot
(with one electron on each dot) system differs from earlier work on
the problem in the sense that it is essentially exact for 
our model.\cite{Harju,Scarola}
Earlier work obtaining the Hilbert space structure of the
coupled double dot quantum computer system used perturbative
approximations akin to Heitler-London or Hund-Mulliken theories.\cite{Loss2,Mizel}  As
explained later in this paper we manage to carry out an exact
numerical diagonalization of the two-electron coupled-dot interaction
Hamiltonian by borrowing theoretical techniques which have been
extremely successful in elucidating the ground and excited state
properties of the strongly correlated fractional quantum Hall system.
Second, we describe and discuss (again using the exact diagonalization
technique as the theoretical tool) a subtle topological feature of a
certain class of solid state spin (``cluster'') qubits, which have
attracted considerable recent attention, where clusters of spins are
cleverly aligned to serve as qubits (rather than just single spins)
offering certain advantages in quantum
computation. \cite{DiVincenzo,Benjamin,Levy,Meier}  We show that in 
looped geometries these
spin cluster qubits have higher-oder spin interaction terms (3-spin,
4-spin, etc. terms) arising from chiral interaction terms which
will have serious adverse consequences for quantum computation by
providing nontrivial flux dependence which must be taken into 
account.\cite{chiral}  Our finding of the chiral Hamiltonian in spin
cluster qubits may have possible consequences for all spin quantum
computing architectures (and not just the semiconductor nanostructure
based ones \cite{Pachos}).  Again, the exact diagonalization technique is the
theoretical method we employ to demonstrate, validate, and quantify
the chiral spin Hamiltonian.  

The common theme running through the two complementary topics studied
in this work is a thorough investigation of the precise applicability
of the Heisenberg Hamiltonian as a description of the spin
interaction in exchange gate quantum computer architectures.  In
particular, we are interested in figuring out the limitations, and
the constraints on the Heisenberg interaction model as the underlying
Hamiltonian for exchange gate, two qubit entanglement in the presence
of an external magnetic field.  In this context we also want to know
what, if any, (and how large) the correction terms are to the
Heisenberg model description of the exchange gate quantum dynamics in
the solid state spin-based quantum computer.  This study is motivated
by the fact that precise knowledge of the exact Hamiltonian
controlling the qubit dynamics is absolutely essential in developing
quantum computing algorithms (and architectures) since in quantum
computation the Hamiltonian itself determines the programming codes.
We must, therefore, know the Hamiltonian precisely.

To study modifications to the Heisenberg Hamiltonian in tunnel coupled 
single-electron quantum dots in an external magnetic field  we 
consider a realistic Hamiltonian involving an equal number of 
parabolic quantum dots and electrons interacting through the Coulomb
interaction.  We find deviations from simple Heisenberg 
behavior where expected, when the quantum dots are strongly coupled 
and under intense magnetic fields.  
We find, using a variational ansatz\cite{Kawamura} and exact diagonalization, 
that at high magnetic fields the 
two-electron orbital states may be characterized by a 
vorticity.  We define vorticity as the 
number of zeroes attached to each electron in the many-body wave
function through the term: $(z_i-z_j)^p$, where $z=x+iy$ 
is a complex coordinate in the
$x-y$ plane and the integer $p$ is the vorticity of the state.  
We find level repulsion between states with either 
even or odd vorticity.  We identify the parameter
regime required to maximize level repulsion among wanted 
and unwanted states ensuring 
adiabatic operation of the exchange gate.   

Using perturbation theory applied to the extended Hubbard model 
we find other qualitative modifications to the
Heisenberg interaction among weakly tunnel-coupled quantum dots.  
We find that in spin clusters formed from loops of three
or more spins, many-spin terms couple to external sources of flux.  
We focus, numerically, on the triangular configuration in
particular.  Here, we show quantitatively that 
a three spin chiral term couples to 
flux through the triangle.\cite{chiral}  
We investigate both the low and high magnetic field regimes and model 
the effective spin Hamiltonian in terms of two parameters: The flux
passing through the triangle {\em and} the vorticity (or effective
flux) attached to each electron.   
We go on to study four spin configurations where 
three and four body terms modify 
the Heisenberg interaction.\cite{Mizel}  We find that four body terms 
depend on the flux through closed loops as well.             

Implementations of exchange based-only 
quantum computation with the least overhead, from a quantum computing
perspective, involve several spins interacting 
simultaneously through the Heisenberg interaction. 
Although, this is not necessary.  One may consider algorithms  
involving no more than two simultaneously coupled spins or geometries
which exclude closed loops of spins.   
In what follows we analyze a special case: looped geometries 
of simultaneously interacting single-spin quantum dots which, 
as we will show, necessarily involve flux dependence.

In Sec. II we present a model of several lateral, single-electron
quantum dots in an external magnetic field.  The model establishes 
four regimes defined by parameters related to confinement, the
external magnetic field, and inter-dot separation.  In Sec. IIA 
we relate the coupled quantum dot model to a fourth order 
spin Hamiltonian based on well known perturbation 
theories of the extended Hubbard model.  Here, we broaden this
treatment to include magnetic field effects.  We find many-body spin terms
which couple to external magnetic fields.  In Sec. IIB we describe 
a variational theory of the many-dot problem which accesses 
non-perturbative regimes of the parameter space.  In Sec. III we 
discuss implications of the perturbative and variational treatments 
to qubit proposals involving two, three, and four coupled spins.  
In Sec. IVA we use exact diagonalization of 
a physically plausible, Coulomb Hamiltonian 
to explore the low energy Hilbert space of two, coupled, single
electron quantum dots.  We find that the variational
ansatz agrees with exact results.  The results suggest 
that electrons in coupled
quantum dots capture vortices of the many-body wave
function to screen the Coulomb interaction, at 
high fields.  In Sec. IVB we compare exact
diagonalization results of three, coupled, single-electron quantum dots 
with the spin Hamiltonian derived using perturbation theory.  We show, 
numerically that chiral, three spin terms couple the spin states of a
decoherence-free subsystem to external sources of enclosed flux.  We 
parameterize the magnetic field behavior of the lowest spin states
with a spin Hamiltonian which depends on the enclosed flux and the 
number of vortices (or effective flux \cite{Kawamura}) attached to each electron.    
We conclude in Sec. V.

\section{Model}

We study the low energy Hilbert space 
of $N$, lateral quantum dots containing $N$ electrons lying in the
$x-y$ plane with the following Hamiltonian:
\begin{eqnarray}
H&=& \sum_{i=1}^{N}\left[ \frac{1}{2m^*}
\left( \textbf p_i+\frac{e}{c}\textbf A_i\right)^2 +V_N(\textbf r_i) \right]
\nonumber
\\ 
&+&\sum_{i<j}^{N} \frac{e^2}{\varepsilon \vert \textbf r_i-\textbf r_j \vert}
+g^*\mu_B \textbf S\cdot\textbf B. 
\label{FullH}
\end{eqnarray}
We focus on GaAs.  In which case we have 
an effective mass $m^*=0.067m_e$, dielectric
constant $\varepsilon=12.4$, and $g$-factor $g^*= -0.44$.  We work in the
symmetric gauge with magnetic field $\bm{B}=B\hat{\bm{z}}$.  
$\textbf{S}$ is the total spin.  The single 
particle potential confines the electrons to
lie in parabolic wells centered at $N$ positions, $\textbf{R}_i$:  
\begin{eqnarray}
V_N(\textbf r)=
\frac{m^*\omega_0^2}{2}\min\{\vert \textbf r-\textbf R_1 \vert^2,...,\vert\textbf r-\textbf R_N \vert^2 \},
\label{pot}
\end{eqnarray}
where $\omega_0$ is a parabolic confinement parameter which may be
compared to the cyclotron frequency, $\omega_c=eB/m^*c$.  The
confining potential localizes the electrons at $N$ sites with
inter-site separation 
$R=\vert \textbf{R}_i-\textbf{R}_j\vert$.   $R$ may be compared to the 
modified magnetic length 
$a=\sqrt{\hbar c/eB}(1+4\omega_0^2/\omega_c^2)^{-\frac{1}{4}}$. 
Solutions of the above 
Hamiltonian take the form $\mathcal{A}[\vert\psi
\rangle\otimes\vert\lambda\rangle_N]$ where $\vert \psi \rangle$ and 
$\vert \lambda \rangle_N$ 
are the orbital and spin parts of the wave function, 
$\mathcal{A}$ is the antisymmetrization operator, and the subscript denotes the
number of electrons and quantum dots.   

The parameters in $H$ define several regimes relevant 
for quantum computing architectures utilizing similarly confined 
single-electron quantum dots in an external magnetic field.
\begin{figure}
\includegraphics[width=2.3in]{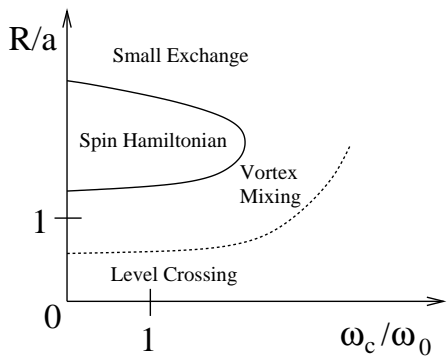}
\caption{ 
Schematic showing four regions 
characterizing spin and orbital degrees of freedom in 
several coupled quantum dots 
defined by the parameters in 
Eq.~(\ref{FullH}) at fixed confinement, $\omega_0$.
The vertical axis depends on the 
inter-dot spacing, $R$, while the horizontal axis depends on the 
ratio between the cyclotron frequency, $\omega_c$, and $\omega_0$.     
The region of interest for quantum computing (spin Hamiltonian) 
yields a spin Hamiltonian dominated by Heisenberg
exchange coupling, $J_{ij}\textbf{S}_i\cdot \textbf{S}_j$.  
Here, the higher orbital energy levels of the quantum dot 
 have much higher energy than the exchange splitting, $\Delta\gg J$.  
Above this region, $R/a\gg 1$, the electrons in each 
dot do not interact strongly.  
At high magnetic fields and near $R/a\sim 1$, 
the electrons capture vortices of the many-body wave function to 
form mixtures with $\Delta\gtrsim J$.  Below the dotted line the 
small separation between dots allows 
single dot behavior, and therefore 
level crossing, $\Delta \lesssim J$.
\label{regime}}
\end{figure}
Fig.~\ref{regime} depicts four separate pieces of the 
parameter space with the confinement parameter, $\omega_0$, fixed.  
The solid line encloses an area in which  
the excited, orbital states of
the quantum dots have high energy, $\Delta$, and the inter-dot
coupling between 
two dots maps onto the Heisenberg Hamiltonian,
$J_{12} \textbf{S}_1 \cdot \textbf{S}_2$, as originally envisaged 
in Ref.~\onlinecite{Loss}.  We show, using both perturbation theory in the
Hubbard limit (Sec. IIA) and 
exact diagonalization (Sec. IVB), that in this regime several coupled 
quantum dots, $N > 2$, involve symmetry breaking many-spin terms with 
nontrivial flux field dependence.  Above the spin Hamiltonian 
regime $(R/a\gg 1)$ weak inter-dot tunneling yields a 
small exchange interaction.  Below the dashed line, higher 
orbital states of the quantum dot have a particularly low 
energy and therefore mix with the low energy 
states of the quantum dot, $\Delta \lesssim J$.  This regime is characterized by 
magnetic field dependent level crossings among potential qubit spin
states and unwanted higher energy levels of the double dot system.   Between 
the level crossing and spin Hamiltonian regimes 
we find, using a combination of exact diagonalization (Sec. IV) and 
variational techniques (Sec. IIB), that electrons in the dots capture vortices 
of the $N$-body wave function to screen the strong Coulomb
interaction.  Here we find that the first excited state of the many dot
system mixes vorticity leaving the unwanted excited states of the quantum 
dot somewhat higher in energy than the spin splitting between the
lowest states, $\Delta \gtrsim J$.  In Sec. II we discuss how these 
regimes pertain to qubit proposals 
utilizing coupled quantum dots.

\subsection{Perturbative Expansion}

We first seek an approximation to Eq.~(\ref{FullH}) that
qualitatively captures the structure of the low energy Hilbert space in terms 
of on-site spin operators in the limit of weak inter-site coupling.
Such a Hamiltonian may be used to define qubit gates among spins 
localized in neighboring quantum dots.
We work in the single band, tight binding limit, a good
approximation in the limit $\Delta \gg J$.  
We also take the on-site Coulomb interaction to be much larger 
than the tunneling energy.  
As a result we may also restrict our attention to singly occupied 
states.  As a first approximation we then obtain the 
extended Hubbard Hamiltonian:
\begin{eqnarray}
H_{\text H}&=&
-\sum_{i,j,\alpha\in
\uparrow\downarrow}t_{ij}c_{i\alpha}^{\dagger} c_{j\alpha}^{\vphantom{\dagger}}
+U\sum_{i}n_{i\uparrow}n_{i\downarrow}
\nonumber
\\
&+&V\sum_{i,(\alpha,\alpha')\in\uparrow\downarrow}n_{i,\alpha}n_{i+1,\alpha'}
+g^*\mu_B\textbf B\cdot\sum_i \textbf S_i, 
\label{hub}
\end{eqnarray}
where $c_{i\alpha}^{\dagger}$ creates a fermion at the site $i$ with
spin $\alpha$ and 
$n_{i\alpha}=c_{i\alpha}^{\dagger} c_{i\alpha}^{\vphantom{\dagger}}$.
$H_{H}$ incorporates the on-site spin operators, 
$\textbf{S}_i=\frac{1}{2}c_{i\alpha}^{\dagger} \bm{\sigma}_{\alpha\alpha'} 
c_{i\alpha'}^{\vphantom{\dagger}}$, where $\bm{\sigma}$ are the Pauli
matrices.  In the presence of an external magnetic field 
the tunneling coefficients 
are complex:
$t_{ij}=\vert t_{ij} \vert \exp (2\pi i\Phi_{ij}/\Phi_0)$,
where $\Phi_0\equiv hc/e$ is the flux quantum.
The magnetic vector potential generates the Peierls phase:
\begin{eqnarray}
\Phi_{ij}=\int_{i}^{j}\textbf A\cdot d\textbf r,
\end{eqnarray}
where the integral runs along a path connecting the sites 
$i$ and $j$.
Working in the limit $\vert t_{ij}\vert/U \ll 1$ and $v\equiv V/U\ll 1$
we can self-consistently confine our attention to the single occupancy states of  
the full Hilbert space.  For large $v$ the extended Hubbard term 
favors double occupancy and our approximation breaks down.  We
consider this regime variationally in the next section.

We expand $H_{\text{H}}$ 
by applying a unitary transformation 
$\exp(iK)H_H\exp(-iK)$, where $K$ is an operator 
changing the number of doubly occupied
states.\cite{Macdonald,Takahashi,vanDongen,Rokhsar,Sen}  We 
obtain, up to constant terms, the following spin Hamiltonian:
\begin{eqnarray}
&H_{\text{eff}}&=g^*\mu_B\textbf B\cdot\sum_{i} \textbf S_i
+\sum_{i,j}\mathcal{A}^{(1)}_{i,j}\textbf{S}_i\cdot \textbf{S}_j
\nonumber
\\
&+&\sum_{i,\tau\neq\pm\tau'}\mathcal{A}^{(2)}_{i,\tau,\tau'}\textbf{S}_{i+\tau}\cdot
\textbf{S}_{i+\tau'}
+\sum_{i,\tau}\mathcal{A}^{(3)}_{i,\tau,-\tau} 
\textbf{S}_{i+\tau}\cdot \textbf{S}_{i-\tau}
\nonumber
\\
&+&\sum_{ijk\in\bigtriangleup}\mathcal{B}_{i,j,k}\textbf{S}_i\cdot
\textbf{S}_j\times \textbf{S}_k 
\nonumber
\\
&+&\sum_{ijkl\in\square}\mathcal{C}_{i,j,k,l}
[(-\textbf{S}_i\cdot \textbf{S}_j+\frac{1}{4})
(-\textbf{S}_k\cdot \textbf{S}_l+\frac{1}{4})
\nonumber
\\
&+&(-\textbf{S}_j\cdot \textbf{S}_k+\frac{1}{4})
(-\textbf{S}_i\cdot \textbf{S}_l+\frac{1}{4})
\nonumber
\\
&-&(-\textbf{S}_i\cdot \textbf{S}_k+\frac{1}{4})
(-\textbf{S}_j\cdot \textbf{S}_l+\frac{1}{4})]
+\vartheta\left(\frac{t^5}{U^4}\right),
\label{eff}
\end{eqnarray}
where we define the following flux independent factors:
\begin{eqnarray}
\mathcal{A}^{(1)}_{i,j}=2\left[\frac{\vert t_{ij}\vert^2}{U(1-v)}
-\frac{4\vert t_{ij}\vert^4 (1+v)}{U^3(1-v)^3}
\right]
\nonumber
\end{eqnarray}
\begin{eqnarray}
\mathcal{A}^{(2)}_{i,j,k}= \frac{-4(1-2v)\vert t_{ij}\vert^2\vert
  t_{ik}\vert^2}{U^3(1-v)^3}
\nonumber
\end{eqnarray}
\begin{eqnarray}
\mathcal{A}^{(3)}_{i,j,k}=\frac{2(1+v)\vert t_{ij}\vert^2\vert
  t_{ik}\vert^2}{U^3(1-v)^3},
\nonumber
\end{eqnarray}
and the following flux dependent factors:
\begin{eqnarray}
\mathcal{B}_{i,j,k}=\frac{24
\vert t_{ij}\vert \vert t_{jk}\vert  \vert t_{ki}\vert}{U^2(1-v)^2}
\text{sin}\left(\frac{2\pi \Phi_{ijk}}{\Phi_0}\right)
\nonumber
\end{eqnarray}
\begin{eqnarray}
\mathcal{C}_{i,j,k,l}=\frac{16(5-12v)
\vert t_{ij}\vert \vert t_{jk}\vert  \vert t_{kl}\vert\vert
t_{li}\vert}
{U^3(1-v)^2(1-3v)}\text{cos}\left(\frac{2\pi \Phi_{ijkl}}{\Phi_0}\right).
\nonumber
\end{eqnarray}
The second sum in Eq.~(\ref{eff}) 
contains the usual Heisenberg term 
if we define $J_{ij} = 2\vert t_{ij}\vert^2/U$.  The third and fourth
sums modify the Heisenberg term and depend on the lattice vectors
$\tau$ connecting a site to its neighbor.  The fifth sum is a three site sum over chiral \cite{Wen} terms around
distinct, closed loops ($\triangle$), 
denoted $\chi_{ijk}=\mathcal{B}_{i,j,k}\textbf{S}_i\cdot
\textbf{S}_j\times \textbf{S}_k $. 
The sixth sum includes four sites 
around distinct, closed loops ($\square$).  
The coefficients in the last two
sums depend on the flux enclosed by three site loops, 
$\Phi_{ijk}$, and four site loops, $\Phi_{ijkl}$.
$H_{\text{eff}}$ applies to any half-filled, single-band, 
singly occupied lattice in the presence of an 
external magnetic field, and in the limits $t/U\ll 1$ and $v\ll 1$. 

The magnetic field dependence of the tunneling matrix elements
may be calculated directly using a Wannier basis
formed from dot-centered, Gaussian, single particle states.  
An explicit form for $\vert t_{12} \vert^2/U$ may be found in 
Ref.~\onlinecite{Loss2} with a slightly different confinement than the
one defined in Eq.~(\ref{pot}).  
Ref.~\onlinecite{Loss2} contains two results relevant for our
discussion.  One first observes that, in the Hubbard approximation, 
$\vert t_{12} \vert^2/U$ decreases exponentially with increasing
magnetic field or inter-dot separation.  Second, one finds 
$J_{12}\sim\vert t_{12} \vert^2/U >0$, 
for all magnetic fields and inter-dot separations.

\begin{figure}
\includegraphics[width=2.3in]{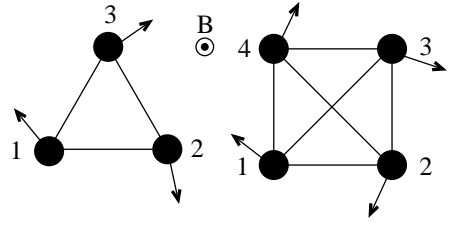}
\caption{ 
Two possible looped configurations of lateral  
quantum dots containng single spins in a 
magnetic field.  
The triangular configuration contains three tunneling 
channels with one loop formed by the vertices $1231$. 
The square configuration contains, in general, six tunneling 
channels with seven disctinct loops formed by 
traversing vertices $1231$, $2342$, $3413$, $4124$, 
$12341$, $12431$, and $14231$. 
\label{dots}}
\end{figure}

Eq.~(\ref{eff}) incorporates three and four body spin terms.
The three body chiral term splits the energy between states involving third order, 
virtual tunneling processes along and counter to the applied vector
potential.  The phase, $2\pi\Phi_{ijk} /\Phi_0$, 
is the Aharonov-Bohm phase 
generated by the virtual current moving around the 
flux enclosed by the three site loop.  
The chiral term vanishes on bipartite lattices as a result of particle hole 
symmetry.\cite{Sen}  It plays a particularly active role in triangular
lattices, Fig.~\ref{dots}.    

While the chiral term vanishes in the zero field limit, 
four body terms survive for $N>3$. 
Four body terms have been discussed in relation to coupled quantum dot
systems in a 
four site, tetrahedral geometry. \cite{Mizel}  
In the tetrahedral geometry there are three terms in the sum over four site 
loops (three distinct $\square$'s)
yielding a particle exchange symmetric, four body 
contribution to the spin Hamiltonian: 
$\sim\sum_{ijkl\in\square}(\textbf{S}_i\cdot \textbf{S}_j)(\textbf{S}_k\cdot \textbf{S}_l)$.  
We may, similarly, consider a 
two dimensional square geometry with equal tunnelling 
between all sites.  Here we find three distinct, four-site loops, 
Fig.~\ref{dots}.  These add to give the term discussed in 
Ref.~\onlinecite{Mizel}.  In the absence of diagonal tunneling only 
the exterior loop, $12341$, survives, yielding the symmetry 
breaking four body term: 
$(\textbf{S}_1\times\textbf{S}_2)\cdot(\textbf{S}_3\times\textbf{S}_4)
+(\textbf{S}_1\cdot\textbf{S}_2)(\textbf{S}_3\cdot\textbf{S}_4)$,
which, through $\mathcal{C}_{ijkl}$, shows flux dependence.   

As the inter-dot separation decreases we leave the single-band,
single-occupancy limit and a real current (rather than a virtual one)
splits the energy spectrum.  The spin splitting
in this limit depends on first order, direct exchange processes rather
than the second order, superexchange processes discussed above.
Our anaylsis based on the extended Hubbard model may 
miss qualitative features in the low energy Hilbert space by poorly estimating 
the coefficients of different spin terms.  Nonetheless, symmetry 
requires that, in the weakly-coupled lattice
limit Eq.~(\ref{eff}) includes all possible 
scalars formed from spin operators, up to and including 
four body terms.  While the magnitude of these coefficient 
may be estimated accruately by expanding the Hilbert space, 
we seek to study the qualtitative physics associated with 
changes in sign.  To go beyond our single band lattice model we
account for qualitative changes in sign by modelling 
orbital effects with a variational ansatz.

\subsection{Variational States}

We now discuss a set of variational states which model  
the low energy orbital states of Eq.~(\ref{FullH}) 
in an effort to go beyond the Hubbard limit discussed above.  
To obtain an accruate variational wavefunction 
we examine the form of the exact wavefunction in two limits:  the
upper left and lower right corners of Fig.~\ref{regime}.  We then 
construct an ansatz which connects both regimes.
We begin with the simplest 
system, $N=2$.  It is analytically soluble in two extreme regimes:  
Two well-separated one-electron ``artificial atoms'' and a 
two-electron artificial atom in a high magnetic field. 
The first case is trivial and consists of two well separated 
quantum dots (akin to two well-separated one electron atoms $not$ 
in a molecular state) with one electron in each dot, $R/a \gg 1$.  In this case we may ignore the Coulomb 
interaction.  The non-interacting ground state consists of 
degenerate singlet and triplet states.

In the second soluble limit (a two-electron artificial atom) 
two electrons lie in one 
parabolic dot in a strong magnetic field.  In this case 
we take $\omega_c/\omega_0 \gg 1$ and $R=0$.  Correspondingly, 
the relative and $z$-component of angular momentum 
commute with the Hamiltonian.
At large magnetic fields we may project Eq.~(\ref{FullH}) onto 
the lowest Landau level (LLL), giving:  
\begin{eqnarray}
H\left\vert_{\omega_0\ll\omega_c,R=0}\right.
=\gamma\hat{L}_z
+\sum_{m=0}^{\infty}\mathcal{V}_m \hat{P}_{m},
\label{LLLH}
\end{eqnarray}
where we define $\gamma\equiv\frac{\hbar}{2}
(\sqrt{\omega_c^2+4\omega_0^2}-\omega_c)$ and  
$\hat{L}_z$ is the total angular momentum    
in the $z$ direction.  The second term represents 
the LLL Coulomb interaction, projected 
onto eigenstates of relative angular momentum, $m$, 
via the projection operator, $\hat{P}_{m}$.  The 
coefficients, $\mathcal{V}_m$, are the Haldane pseudopotentials 
\cite{Haldane} which, for the Coulomb interaction, 
decrease with increasing $m$, at large $m$.  The unnormalized 
eigenstates of relative angular momentum are:
\begin{eqnarray}
\vert m\rangle=(z_1-z_2)^{m}
\exp\left(\frac{-\vert z_1\vert^2-\vert z_2\vert^2}{4a^2}\right),
\label{Rzstates}
\end{eqnarray}
where $z=x+iy$.  It can be shown directly 
that, because there is no center of mass motion, 
the above wave functions are also eigenstates 
of $\hat{L}_z$, with eigenvalue $m$.  Thus the 
set of states $\vert m\rangle$ form 
an orthogonal set of eigenstates of 
Eq.~(\ref{LLLH}), with eigenvalues 
$E_m=\gamma m+ \mathcal{V}_m$.  The relative angular momentum 
of the lowest energy state depends on the parameters in 
$\gamma$ and the form of the interaction.  For the 
LLL Coulomb interaction,
in the artificial zero field limit, the lowest energy 
state has $m=1$.  Increasing $B$ lowers the confinement 
energy cost, $\gamma m\sim m/B$, and raises 
the Coulomb cost, $\mathcal{V}_m \sim \sqrt{B/(m+1)}$, thereby raising $m$ by one.  
The transition from one eigenstate to the next occurs 
when $E_m=E_{m+1}$ which, for $\omega_c/\omega_0\gg 1$, occurs 
at magnetic fields:
\begin{eqnarray} 
B_m\approx\left(\frac{C}{\tilde{\mathcal{V}}_{m}-\tilde{\mathcal{V}}_{m+1}}
\right)^{\frac{2}{3}},
\label{fields}
\end{eqnarray}
where $\tilde{\mathcal{V}}_{m}\equiv \mathcal{V}_m/(e^2/4\pi\varepsilon a)$ and 
$C\equiv4\pi\varepsilon\hbar^{3/2}\omega_0^2 m^*  e^{-7/2}$.
For $\hbar\omega_0=3$ meV we find $C[\text{T}^{3/2}]\sim 1.2$.
The states, $\vert m \rangle$, are symmetric (antisymmetric) 
with respect to particle exchange if $m$ is even (odd).  
The total wave function, $\mathcal{A}[\vert\psi\rangle\otimes\vert\lambda\rangle_2]$, must 
be antisymmetric.  Therefore $\vert\lambda\rangle_2$ is spin singlet (triplet) 
for $m$ even (odd).  Here, the index $m$ may be interpreted 
as the number of zeros or vortices attached to each electron, 
allowing us to assign a vorticity to each spin state. 

We now construct variational states which reproduce the exact results 
discussed above and the low energy physics of the
intermediate, physical parameter regime as well, $R/a\sim 1$ and
$\omega_c/\omega_0 \sim 1$.  
The composite fermion theory \cite{Jain} of the 
fractional quantum Hall effect offers an accurate variational ansatz
describing two dimensional electron systems at high magnetic fields.
A composite fermion is the bound state of an electron and an even
number of quantum mechanical vortices of the many-body wave 
function.  The corresponding 
orbital wave function is \cite{Jain,Kawamura}: $\psi=\mathcal{J}\phi$, where 
$\phi$ is a weakly interacting fermion state, $\mathcal{J}$ a Jastrow factor,
and $\psi$ the highly correlated state of electrons.  
In isotropic, spinless systems $\mathcal{J}$ attaches 
an even number of vortices to the fermions in the antisymmetric 
state $\phi$ yielding an antisymmetric electron 
wave function.  In anisotropic systems with 
additional quantum numbers one may bind 
an even $or$ odd number of vortices to 
each particle while preserving the antisymmetry of the 
overall wave function.\cite{Fertig,Scarola2}
Applying the composite fermion ansatz to the Hamiltonian studied 
here we take $\phi$ to 
be the non-interacting ground state of Eq.~(\ref{FullH}).  A more
rigorous, but technically demanding, approach for large systems 
should use the weakly interacting ground state instead.    
We also 
take $\mathcal{J}=\prod_{j<k}(z_j-z_k)^{p}$, where $p=0,1,2,..$, giving:
\begin{eqnarray}  
\overline{\psi}_p=\prod_{j<k}(z_j-z_k)^{p}\phi.
\label{states}
\end{eqnarray}
This is our initial, high field solution of  
Eq.~(\ref{FullH}) at arbitrary $N$.  $\overline{\psi}_p$ reduces 
to Eq.~(\ref{Rzstates}) at $R=0$.  For $R/a\gg 1$ the fermions in the 
state $\phi$ become localized on each dot leaving  
$\mathcal{J}$ constant $\sim R^p$.  In which case 
$\overline{\psi}_p$ reduces to the limit of 
two independent electrons.  

To improve the variational states in the low field 
regime, $\omega_c/\omega_0 \leq 1$, we note that the 
Coulomb energy cost may be lowered by mixing with higher 
energy states of the quantum dot thereby increasing the 
average inter-electron separation.  In the 
$R=0$ limit, rotational symmetry requires the addition  
of states with the same angular momentum.
This leads to the following trial states:
\begin{eqnarray} 
\psi_p=\prod_{j<k}(z_j-z_k)^{p}\left(1
+\beta b^\dagger a^\dagger \right)\phi,
\label{general}
\end{eqnarray}
where the variational parameter $\beta$ 
controls the amount of mixing with higher 
energy levels of the quantum dot.
The total raising operators $b^\dagger=b^\dagger_1+...+b^\dagger_N$ and 
$a^\dagger=a^\dagger_1+...+a^\dagger_N$ act on the 
Fock-Darwin basis states centered between the 
dots.  The single particle 
raising operators are given by: 
$b_j^\dagger=(z_j^*/2-2\partial_{z_j})/\sqrt{2}$
and
$a^\dagger_j=i(z_j/2-2\partial_{z_j^*})/\sqrt{2}$.
The above variational states include mixing with 
higher energy states of the same angular momentum because the 
operator $b^\dagger a^\dagger$ does not change the angular momentum 
of a Fock-Darwin state.  They will be tested in Section IV.

\section{Spin Based Quantum Dot-Quantum Bits}

Gate operations on single and multi-spin qubits 
rely on the adiabatic evolution of the spin state under the 
unitary time evolution operator defined in 
terms of the appropriate spin Hamiltonian ({\em e.g.} Eq.~(\ref{eff})):
\begin{eqnarray}  
\mathcal{T}\text{exp}\left(-i\int_{0}^{T}H_{\text{eff}}(t)dt\right)\vert \lambda(t'=0)\rangle_N,
\end{eqnarray}
where $\mathcal{T}$ indicates time ordering and $T$ the duration of a
gate pulse.  The qualitative spin physics captured by the spin Hamiltonian, 
Eq.~(\ref{eff}), therefore plays a crucial role in defining gates
formed from coupled quantum dots.  Concurrently, the orbital states, 
Eq.~(\ref{general}), can be used to calculate the parameters in 
Eq.~(\ref{eff}) and their regime of applicability.  
In forming quantum gates out of coupled, single
spin quantum dots we study 
modifications to the Heisenberg paradigm and its implications to 
qubit proposals in three 
different systems:  
A)  Two strongly coupled quantum dots.  
B)  Three simultaneously, weakly coupled quantum dots.  
C)  Four simultaneously, weakly coupled quantum dots.    

\subsection{Two Quantum Dots}

Solid state qubit proposals often make use of the 
Heisenberg exchange interaction between spins in neighboring quantum dots.
The exchange interaction offers the
potential for a universal set of quantum gates through the adiabatic
operation of the exchange gate with \cite{Loss} or without
\cite{DiVincenzo} single spin operations.  Application of the exchange gate 
to the two dot system will 
be an adiabatic process if the energy between the lowest, 
unwanted excited state of the 
double quantum dot and the highest spin state storing quantum
information is much larger than the exchange splitting, $\Delta\gg
J_{ij}$.  This condition 
is satisfied in the spin Hamiltonian regime in Fig.~\ref{regime}   
but is not necessarily met if we change 
the parameters in Eq.~(\ref{FullH}) slightly because $J$ has
exponential $B$ and $R$ dependence at 
large $B$ and $R$, respectively.\cite{Loss2}  
In fact, experiments on 
coupled quantum dots, while pushing for shorter gate times 
(and hence larger exchange energies), may indeed 
leave the border defined by the solid line in 
Fig.~\ref{regime}.\cite{Loss2,Hu_leak}  It is therefore important to understand 
the low energy Hilbert space of the coupled dot system when $J_{ij}\gtrsim
\Delta$.  We will show, for $N=2$, that the variational states discussed in
Sec. IIB capture the magnetic field dependence of $\Delta$. 
At large fields the variational states describe a bound state between 
electrons and vortices of the $N$-body wave function.  We find the smallest 
$\Delta$, $\Delta_{\text{min}}$, to occur when the 
vorticity (the number of vortices attached to each electron) 
of the first excited states mix to form an anti-crossing. 

\subsection{Three Quantum Dots}

An accurate characterization of the double dot system 
allows us to define the appropriate 
parameter regime in which to study several Heisenberg coupled 
quantum dots and associated magnetic field effects.  Two spin states 
of a three quantum dot structure can serve as an encoded qubit. 
We first construct encodings which protect quantum information 
stored in many-body spin states.  These encodings assume a 
noise operator which, as a demonstration, we choose to be 
collective or Zeeman-like.  
We then search for degeneracies in 
the set of states generated by these noise operators.   
The $S=1/2$ sector of the $N=3$ system provides a 
simple example of a quantum dot, 
decoherence-free subsystem \cite{Viola,Viola2}.
Consider three electrons confined to three quantum dots whose 
centers lie at the vertices of an equilateral triangle 
as shown schematically in Fig.~\ref{dots}. 
In this case a decoherence-free subsystem makes use of a fourfold degeneracy 
at $B=0$ to protect quantum information stored in the
qubit defined by $\vert\lambda\rangle_{3}$, where 
$\lambda=0$ or $1$, from fluctuations in the Zeeman energy.         
The four states are\cite{Nakamura}:
\begin{eqnarray}
\vert\lambda\rangle_{3} &\otimes&\vert -1/2\rangle
\nonumber
\\
&=& \frac{-1}{\sqrt{3}}
\left( \vert \downarrow \downarrow \uparrow \rangle 
+\omega^{\lambda +1}\vert\downarrow\uparrow\downarrow\rangle
+\omega^{2-\lambda }\vert\uparrow\downarrow\downarrow\rangle\right)  
\nonumber
\\
\vert\lambda\rangle_{3}&\otimes&\vert +1/2\rangle
\nonumber
\\
&=& \frac{1}{\sqrt{3}}
\left( \vert \uparrow \uparrow \downarrow \rangle 
+\omega^{\lambda +1}\vert\uparrow\downarrow\uparrow\rangle
+\omega^{2-\lambda }\vert\downarrow\uparrow\uparrow\rangle\right),  
\label{Nthree}
\end{eqnarray}
where $\omega\equiv \exp(2\pi i/3)$.  The second term in the 
tensor product denotes the total $z$-component of spin.   

Up to second order, Eq.~(\ref{eff}) allows an  
encoding against Zeeman-like or collective noise.  
By collective noise we mean an interaction between spins and the 
environment which acts the same on all spins.
By construction, the Zeeman term may alter the energy difference between 
the states in Eq.~(\ref{Nthree}) with different $S_z$, but not 
$\lambda$.  However, the chiral term 
in Eq.~(\ref{eff}) acts {\em non-collectively}.  The chiral 
and Zeeman terms remove 
all degeneracies required to construct a qubit immune to fluctuations
in the perpendicular magnetic field.  Explicitly:    
\begin{eqnarray}
\chi_{123}\vert\lambda\rangle_{3} 
=\frac{\mathcal{B}_{123}}{4}(2\lambda-1)\sqrt{3}\vert\lambda\rangle_{3},
\label{split}
\end{eqnarray}
where $\mathcal{B}_{123}=(12tJ/U)
\text{sin}\left(2\pi\Phi_{123}/\Phi_0\right)$ in the case $J_{ij}=J$
and $\vert t_{ij} \vert=t$ for all $i$ and $j$.

Following Ref.~\onlinecite{Kempe} we may now, 
using Eq.~(\ref{eff}), construct a  
projected spin Hamiltonian which acts on the encoded basis states
$\vert\lambda\rangle_{3}$:
\begin{eqnarray}
\bar{H}_{N=3}=\textbf{F}_3(\mathcal{A}^{(1)},\mathcal{A}^{(2)},\mathcal{A}^{(3)})\cdot\bar{\textbf{S}}
+\frac{\sqrt{3}}{2}\mathcal{B}_{123}\bar{S}_z,
\label{eff3}
\end{eqnarray}
where $\bar{\textbf{S}}$ is a pseudospin operator defined by
projection onto two encoded basis states, $\vert\lambda\rangle_N$ in our case.
$\textbf{F}_N$ is a basis dependent, effective magnetic field 
which may be tuned
through suitable manipulation of $t_{ij}$ and depends {\em only} on the
coefficients of the two body terms in Eq.~(\ref{eff}).  $\textbf{F}_N$ may be 
calculated from these two body terms using the exchange operator: 
$E_{ij}=(4\textbf{S}_i\cdot\textbf{S}_j+I_{ij})/2$, where $I$ is the
identity operator. 
As apparent from Eq.~(\ref{eff3}),  
$\chi_{123}$ yields an {\em effective} Zeeman splitting  
between the encoded basis states of the three spin qubit.  
In Sec. IVB 
we verify numerically that the chiral term is actually sizable 
in the spin Hamiltonian regime of Fig.~\ref{regime}.  We therefore arrive
at a revealing inconsistency in seeking a 
decoherence free subsystem from a looped, three spin system.  Part of our motivation for
simultaneously coupling three spins was to remove the Zeeman 
term as a potential noise source.  However, we have only 
enhanced the system's dependence on the 
external magnetic field by coupling the three spins in a loop. 

In the event that we
may control the flux through the three spin system, the chiral term offers 
an additional one qubit gate.  This term yields two advantages.  The
first stems from a comparison with single spin operations using
localized magnetic fields.  The three spin object encompasses a larger
area than a single spin  
and therefore eases constraints on locally applied 
magnetic fields used in applying single spin gates.  
Second, exchange-only encoded universality schemes 
require a large overhead and extremely accurate application 
of the exchange gate to implement elementary algorithms.  The 
chiral term may offer some relief form these constraints using 
algorithms which include the new, encoded Pauli-Z gate in Eq.~(\ref{eff3}).

\subsection{Four Quantum Dots}

We now turn to the case of four coupled spins, 
the lowest number of physical  
spins supporting a decoherence free subspace.\cite{Whaley,Bacon}
We begin with 
four quantum dots containing four electrons 
coupled with equal tunneling $\vert t_{ij} \vert=t$, including 
diagonal terms.  Fig.~\ref{dots} shows a two-dimensional schematic.   
With equal tunneling we find a decoherence free subspace among 
two $S=0$ states corresponding to 
$\lambda=0$ and $1$:
\begin{eqnarray}
&&\vert\lambda\rangle_{4}=\vert\uparrow\uparrow\downarrow\downarrow\rangle
\nonumber
\\
&+&\vert\downarrow\downarrow\uparrow\uparrow\rangle
+\omega^{\lambda+1}\vert\uparrow\downarrow\uparrow\downarrow\rangle
+\omega^{\lambda+1}\vert\downarrow\uparrow\downarrow\uparrow\rangle
\nonumber
\\
&+&\omega^{2-\lambda}\vert\downarrow\uparrow\uparrow\downarrow\rangle
+\omega^{2-\lambda}\vert\uparrow\downarrow\downarrow\uparrow\rangle.
\label{Nfour}
\end{eqnarray}
Including all single and two-body spin terms
in Eq.~(\ref{eff}), these states show no explicit magnetic field dependence  
(excluding the magnetic field dependence of $\vert t_{ij} \vert$
discussed in Sec. IIA).  
As for the three-body term, the spin Hamiltonian must respect 
the inter-site exchange symmetry inherent in the lattice.  
In the basis $\vert\lambda\rangle_{4}$ we find:  
\begin{eqnarray}
&&\sum_{ijk\in\triangle
} \chi_{ijk}\otimes I_l \vert\lambda\rangle_4
\nonumber
\\
&=&\frac{\sqrt{3}}{4}(2\lambda-1)\sum_{ijk\in\triangle}\mathcal{B}_{ijk}
\epsilon_{ijkl}
\vert\lambda\rangle_4,
\label{chiralsum}
\end{eqnarray}
where $\epsilon_{ijkl}$ is the four component Levi-Civita symbol and the sum 
excludes $l= i,j$ or $k$.  As expected, the sum vanishes 
with tunneling $\vert t_{ij}\vert
=t$ for all $i$ and $j$ even in a uniform, external magnetic field.  

Four spin terms have a simple representation in the
$\vert\lambda\rangle_4$ basis.
The last sum in Eq.~(\ref{eff}), in the case $\vert t_{ij}\vert=t$, 
involves three sums over four site loops. 
Writing the four spin terms with the exchange operator,  
we find that they
act as the identity operator in the basis defined by 
$\vert\lambda\rangle_4$.  
In this case we have a simple, projected 
Hamiltonian:
\begin{eqnarray}
\bar{H}_{N=4}=\textbf{F}_4(\mathcal{A}^{(1)},\mathcal{A}^{(2)},\mathcal{A}^{(3)})\cdot\bar{\textbf{S}}.
\end{eqnarray}
It is important to note that $\textbf{F}_4$ depends only on 
coefficients from two body spin terms of the 
form $\textbf{S}_i\cdot\textbf{S}_j$.  The three 
{\em and} four body spin terms 
(and therefore external sources of flux through closed loops) 
do not affect the symmetric four dot structure with diagonal
tunneling.  They must maintain 
inter-site exchange symmetry imposed by the lattice, 
precisely the symmetry exploited in constructing 
the decoherence free subspace, $\vert\lambda\rangle_4$.

We now consider symmetry breaking effects.  In the absence of diagonal
tunneling ($t_{i,i+2}=0$) only the external loop, around vertices 
 $12341$ in Fig.~\ref{dots}, in the last sum of Eq.~(\ref{eff}) survives.  
The external loop alone breaks particle exchange symmetry.  
The additional term can be written: $(\textbf{S}_1\times\textbf{S}_2)\cdot(\textbf{S}_3\times\textbf{S}_4)
+(\textbf{S}_1\cdot\textbf{S}_2)(\textbf{S}_3\cdot\textbf{S}_4)$,
excluding two body spin terms.
The entire looped term, including two body spin terms, 
contributes the following term to  $\bar{H}_{N=4}$:
$-\mathcal{C}_{1234}(\bar{S}_x+\sqrt{3}\bar{S}_y)$.  
From this term we see that in the square geometry fourth order 
terms not only modify the Heisenberg interaction, 
and hence the effective magnetic field $\textbf{F}_4$, but also add    
and effective in-plane field in the $\vert\lambda\rangle_{4}$ basis.
The size of this effective, in-plane field depends 
on the real, external flux piercing the square plaquette through 
$\mathcal{C}_{1234}\sim \text{cos}(2\pi\Phi_{1234}/\Phi_0)$.

\begin{figure}
\includegraphics[width=2.0in,angle=-90]{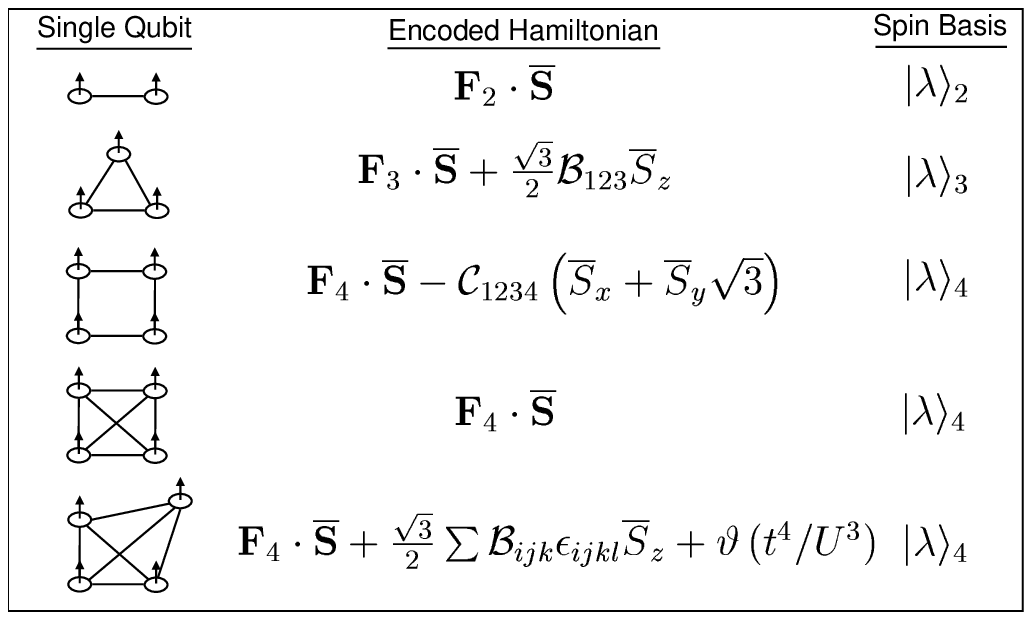}
\caption{ 
Table showing possible multi-spin 
qubits in the left column, the encoded Hamiltonian in the center
column and the corresponding two-state basis in the right column.  
In the left column arrows denote single spins while lines indicate 
tunneling channels which are all equal except in the last 
row.  The Hamiltonians in the center column are written
in terms of the encoded spin $\bar{\textbf{S}}$ defined via the 
two encoded basis states $\vert \lambda\rangle_N$.  For $N=2$ the basis
states are the $S_z=0$, singlet and triplet states while for 
$N=3$ and $4$, $\vert \lambda\rangle_N$ is defined in 
Eqs.~(\ref{Nthree}) and (\ref{Nfour}), respectively.  The effective
magnetic fields ${\bf F}_N$ depend only on the coefficients of two-spin 
Heisenberg terms, $\mathcal{A}^{(i)}$, in Eq.~(\ref{eff}).  The factors 
$\mathcal{B}$ and $\mathcal{C}$
depend on the coefficients of three and four 
spin terms in Eq.~(\ref{eff}) and therefore the flux through closed loops.   
}
\label{table}
\end{figure}

Additional symmetry breaking occurs during 
gate pulses crucial to encoded universality schemes.  In 
order to implement Pauli gating sequences 
on the encoded four spin qubit we must tune 
$\textbf{F}_4$ and therefore the tunneling matrix elements $t_{ij}$.  
When applied to a decoherence free subspace an encoded Pauli gate composed of 
Heisenberg terms must, by construction, involve 
a spin specific asymmetry.
An example was considered in Ref.~\onlinecite{chiral}:
$\vert t_{31} \vert=\vert t_{23} \vert = \vert t_{34} \vert 
= t(1+\delta) $, 
where $\delta$ is a number and all other $\vert t_{ij} \vert=t$.   
The sum over chiral terms in Eq.~(\ref{chiralsum}) then induces an 
energy splitting $24\pi\sqrt{3} t J\delta A B_z /(U\Phi_0)$ between 
the states with $\lambda=0$ and $1$ for $\Phi_{ijk}/\Phi_0\ll 1$.  
Here $A$ is the area of the triangle defined by the vertices $123$ 
in the square geometry of Fig.~\ref{dots}.    
This configuration is depicted in the last row of Fig.~\ref{table}.  
The table summarizes five spin cluster qubit configurations and 
their encoded Hamiltonians written in the $\vert \lambda\rangle_N$
basis.  From the table we see that configurations which break 
inter-site symmetry (rows two, three, and five) 
have non-Heisenberg terms which depend on the flux through closed loops. 

\section{Numerical Results and Discussion}

The accuracy of our perturbative and variational analyses 
may be checked numerically.  We study two systems in particular, 
two electrons in two, adjacent quantum dots and 
three electrons in three adjacent quantum dots arranged in a 
triangle.  We diagonalize the full Hamiltonian, Eq.~(\ref{FullH}), 
in several regimes, including $R/a\sim 1$ and $\omega_c/\omega_0 
\gtrsim 1$.  
We construct the matrix
representing $H$ in the Fock-Darwin \cite{FockDarwin} basis 
centered between the dots.  Previous studies have employed diagonalization of 
similar Hamiltonians using several dot centered basis states.  This 
technique requires lengthy, numerical routines to 
generate an orthogonal set of Wannier basis states.\cite{Hu2,Loss2} 
The limited number of Wannier basis states allows for high accuracy 
only in a regime where
the Coulomb interaction may be treated perturbatively.  However, 
in our treatment we are able to access the strongly correlated regime 
by including up to $\sim 10^5$ Fock-Darwin 
basis states with $z$-component of angular momentum 
less than twelve.  We use 
a modified Lanczos routine to obtain the ground and excited states. 
This technique yields the $entire$ spectrum.  However, here we focus 
on the lowest energy states. 
The energies for $N=2$ converge to within $1 \mu$eV upon 
inclusion of more basis states and may therefore be considered exact. 
While, for the $N=3$ system, the ground and excited state energies converge 
to within $6 \mu$eV upon inclusion of more basis states 
giving a strict variational 
bound to the accuracy.  However, the slow convergence is due to corrections 
in the overall confinement energy cost, 
$\sim 10$ meV.  The energy differences 
quoted here converge much faster 
($< 2$ $\mu$eV) as we increase the number of basis states 
and may therefore be considered exact, with a few exceptions.  These
exceptions occur near degeneracy points where our Lanczos routine 
requires a prohibitive number of steps to discern between two states 
whose energies are to within $5\mu$eV of each other.  In these rare,
but important, cases we extrapolate between the nearest 
convergent energies.

\subsection{Two Quantum Dots}

We seek a quantitatively accurate description of 
the boundaries and underlying physics of all regions 
depicted in Fig.~\ref{regime}.  While we find that 
the perturbative expansion in Sec. IIA is 
valid for $R/a > 1$ and $\omega_c/\omega_0 \lesssim 3$, 
the remaining portions of the parameter space involve long range
correlations.  
Using the
$N=2$ system we check the accuracy 
of the variational ansatz discussed in Sec. IIB in several limits.  We 
expect that the variational states discussed there should 
remain valid for $N>2$, with appropriate modifications. 

\begin{figure}
\includegraphics[width=2.6in]{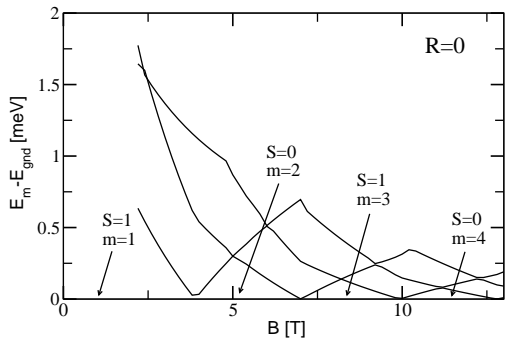}
\caption{
Energy of the 
four lowest states of a single quantum dot ($R=0$) with two electrons 
under a perpendicular magnetic field in the lowest Landau level plotted 
as a function of perpendicular magnetic field.  
The ground state energy is set 
to zero and the parabolic
confinement parameter is $\hbar\omega_0=3$ meV.  
The ground state, with orbital wave function given by
Eq.~(\ref{Rzstates}), alternates between spin singlet $(S=0)$ and 
triplet $(S=1)$ as a function of magnetic field.  The spin singlet and 
triplet states correspond to even and odd angular momentum 
quantum numbers, $m$, respectively. 
\label{lll}}
\end{figure}

We begin with the $R=0$, lowest Landau level limit discussed 
at the beginning of Sec. IIB.  The LLL approximation 
cleanly brings out the physics behind the high field spin transitions in 
two-electron quantum dots but, as we have discussed, needs 
modification at low magnetic fields.  Fig.~\ref{lll}  plots  
$E_m-E_{\textrm{gnd}}$ versus B for the 
four lowest energy states, $m=1,2,3,$ and $4$ with $S_z=0$.  The 
parabolic confinement parameter is $\hbar\omega_0=3$ meV.
Cusps appear 
at $E_m-E_{\textrm{gnd}}=0$ where the ground 
state changes at $B_m$ signaling a change in the number of 
vortices per electron.  (Note that the relation for $B_m$, 
Eq.~(\ref{fields}), is valid for 
$\omega_c/\omega_0 \gg 1$.) 
The ground state clearly shows a number of 
spin transitions with increasing magnetic field.\cite{Wagner}  
Furthermore, the second highest excited state 
becomes degenerate with the third at level crossings which 
occur at magnetic fields between ground state transitions. 
This suggests that quantum information stored in the two lowest energy 
spin states in neighboring quantum dots becomes susceptible to leakage 
when the dots are brought very close together.

\begin{figure}
\includegraphics[width=2.3in]{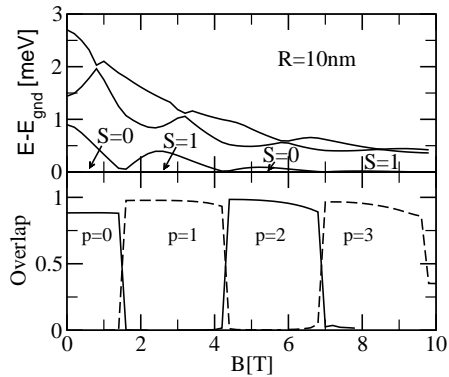}
\caption{
The top panel plots the energy of the four lowest states 
of Eq.~(\ref{FullH}) obtained by exact diagonalization  as a function 
of magnetic field with the ground state energy set to zero.
The separation between parabolic dots is now $R=10$ nm.  The 
parabolic confinement parameters 
is $\hbar\omega_0=3$ meV.  Transitions between 
spin singlet and triplet states remain.  The bottom panel shows the overlap of 
the exact ground state and the trial states given by
Eq.~(\ref{states}).  The number of vortices attached to each electron 
increases with magnetic field from $p=0$ to $p=3$.  As in
Fig.~\ref{lll}, singlet (triplet) states correspond to even (odd) 
values of $p$.  
\label{two_R10}}
\end{figure}

We now turn to the case with finite inter-dot separation, $R>0$ 
outside of the LLL.
The top panel in Fig.~\ref{two_R10} shows the  
four lowest energies obtained from exact diagonalization of 
Eq.~(\ref{FullH}) versus magnetic field.  The energy zero is taken to
be the ground state.  We have chosen an inter-dot separation of
$R=10$ nm, confinement $\hbar\omega_0=3$ meV, and $S_z=0$.  
The energy of the 
first excited state gives the effective exchange splitting  
which changes sign through successive spin transitions 
at each cusp.  The results are qualitatively similar to the results 
shown in Fig.~\ref{lll} but are entirely unexpected.  
Vortex attachment non-perturbatively lowers the Coulomb energy 
of uniform states but does not necessarily apply to 
highly disordered systems.  Yet, the intriguing oscillations in 
the effective 
exchange interaction seen in Fig.~\ref{two_R10} suggest just this 
and therefore require further study.

In comparing Figs.~\ref{lll} and ~\ref{two_R10} we find further 
differences. 
At low fields, the top panel of Fig.~\ref{two_R10} correctly shows 
a spin singlet ground state 
at $B=0$ rather than a triplet state as shown in the unphysical, LLL 
limit of Fig.~\ref{lll}.  Most importantly, the 
degeneracies in excited states at $B=0, 2.4, 5.2$ and $8$ T 
begin to lift, giving $\Delta_{\text{min}} > J$.  
As opposed to the level crossing in 
the single dot, $R=0$ case discussed earlier, 
the breaking of rotational symmetry forces an
anti-crossing among the first and second excited states.  
At small to intermediated inter-dot separations, $R/a\lesssim 1$,  
the higher excited states 
are perturbed, single dot states with a nearly uniform charge density.

A large anti-crossing among the two lowest excited 
states protects the quantum 
information stored in the entangled state of two strongly coupled 
quantum dots.  Experimental uncertainties in $R$ and $\omega_0$ may  
eventually lead to the strongly coupled regime.  Careful study 
of the states making up the anti-crossing is therefore crucial.
The bottom panel of Fig.~\ref{two_R10} plots the overlap of the 
exact ground state and the variational state, Eq.~(\ref{states}), at $R=10$ nm.  
Triplet (singlet) spin states correspond to odd (even) values of 
$p$, as in the $R=0$ case.  The overlaps drop to zero when the 
particle exchange symmetry of the orbital wave function changes. 
We have checked by direct calculation of the density that, 
by $B\sim 9$ T, the modified magnetic length has become small 
enough to localize the electrons on each dot. 
The surprisingly high overlaps prove that vortex
attachment is a valid ansatz even in the highly localized 
regime.  At large dot separations, $R\geq 40$ nm, the Coulomb
interaction lowers to a point where the splitting 
between spin states is near zero at large B.  However, we 
have checked that even here the 
overlaps remain large.  
Another important feature of Eq.~(\ref{states}) 
is that the $p=0$ state does not take into 
account the Coulomb interaction.  The 
overlaps near $B=0$ are correspondingly lower.  

\begin{figure}
\includegraphics[width=2.3in]{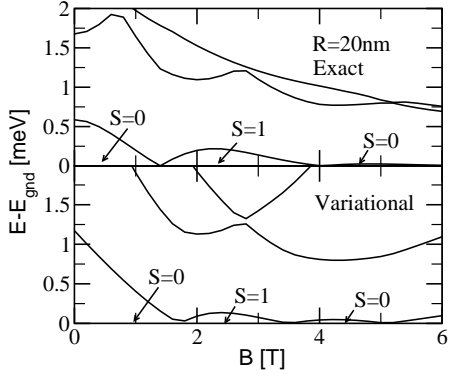}
\caption{  
The top panel shows the same as Fig.~\ref{two_R10} but for a dot
separation of $R=20$ nm.  The bottom panel shows the energy of 
the trial states in Eq.~(\ref{general}) 
as a function of magnetic field.  The ground state energy 
is set to zero.  The energies are obtained by 
orthogonalizing the four modified variational states with 
$p=0, 1, 2,$ and $3$ and diagonalizing Eq.~(\ref{FullH}).  The variational 
parameter $\beta$ is chosen to minimize the total energy.  
\label{two_R20}}
\end{figure}
The top panel in Fig.~\ref{two_R20} plots the exact energy spectrum, as in
Fig.~\ref{two_R10}, but for $R=20$ nm.  Here we see that, at large
magnetic fields, the large separation between electrons localized on 
each dot suppresses the exchange splitting.  However, several 
spin transitions still remain.  The bottom panel in Fig.~\ref{two_R20}
shows the energy of the four variational 
states, Eq.~(\ref{general}), with $p=0, 1, 2,$ and $3$.  We take the 
ground state to be the zero in energy.  We
obtain the energy by orthogonalizing the four 
variational states and diagonalizing Eq.~(\ref{FullH}) in 
this four state basis.  
These variational states are an improvement over Eq.~(\ref{states}).  They 
include mixing with higher energy levels of the dots.  The mixing 
is tuned with the variational parameter $\beta$.
We minimize the energy with respect to $\beta$ at each $B$. 
The parameter 
$\beta$, of the ground state varies from 0.02 at $B=0$ 
to 0.0006 at $B=5$ T showing that 
large magnetic fields all but suppress Landau level mixing.  
The exchange splitting obtained 
with the variational states compares well with the exact value.  
Furthermore, in the range $B=1$ to $4$ T, the second excited state 
captures the essential features of the corresponding exact results.
Rotational symmetry breaking forces the higher excited states 
to open an anti-crossing observed near 
$B=0,2.4,$ and $4.3$ T.
The states at the anti-crossings in Fig.~\ref{two_R20} are 
similar to the states making up the level crossings 
in Fig.~\ref{lll}.  For example the electrons in 
the first excited state at $2.4$ T in Fig.~\ref{two_R20} 
form a two and zero vortex mixed state in a   
56\% to 44\% ratio, as opposed to the ground state 
which holds one vortex per electron, to within 98\%.
To evaluate the anti-crossing 
explicitly we note that for 
$R/a \ll 1$ the asymmetry in confinement acts as a perturbation.  We 
may rewrite the confinement potential up to an overall constant:
\begin{eqnarray}
V_{N=2}(\textbf{r})=\frac{m^*\omega_0^2}{2}\left(\vert\textbf{r}\vert^2-\vert x\vert R\right).
\end{eqnarray}  
The second term breaks rotational symmetry and forces an anti-crossing 
among the lowest two excited states.  It is important to note that the 
two lowest excited states involve states of even vorticity.  
Symmetry allows these two states to mix yielding an 
anti-crossing as one may find by diagonalizing the 
rotational symmetry breaking term in the even-vorticity 
subspace.  The matrix elements are:
$m^*\omega_0^2R/2\left\langle\psi_{p'}\vert \vert 
x_1 \vert + \vert 
x_2 \vert \vert \psi_{p}\right\rangle$, where, 
near $B=2.4$ T for example, 
$p$ and $p'$ may be 0 or 2.  These matrix elements give an 
anti-crossing $\Delta_{\text{min}} \sim m^*\omega_0^2Ra+J$.  This is 
in contrast to ground state transitions between 
states with even and odd vorticity.  Here the states $\psi_p$ and 
$\psi_{p+1}$ cannot mix, allowing the exchange splitting to 
change sign.

We stress that the top panel in Fig.~\ref{two_R20} is obtained 
by diagonalization of the Eq.~(\ref{FullH}) with $\sim 10^5$ basis states while the lower panel is 
obtained by the same method but with four, physically relevant  
basis states.  The 
agreement breaks down at larger fields, $B\sim 5.6$ T, 
because we have not included the $p=4$ variational state in 
the excited states.  Inclusion 
of variational states with large $p$ is necessary at larger fields.  
The excellent agreement obtained thus far demonstrates that the
plethora of spin transitions in strongly coupled double quantum 
dots originates from a swapping of the particle exchange symmetry associated 
with vortex attachment.

\begin{figure}
\includegraphics[width=2.3in]{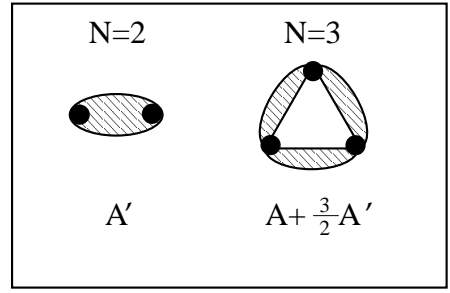}
\caption{  
Schematic diagram showing the area enclosing flux in the 
$N=2$ and $N=3$ systems defined by the density of the $N$-body ground 
state.  The dots represent the centers of the
parabolic quantum dots.  The patterned region for the 
$N=2$ system defines an area $A'$ symmetric about the axis joining the
two dots.  For the $N=3$ system 
the triangular region encloses an area $A$.  A correction to the
triangular region, shown by the 
three patterned additions, defines an area $\approx \frac{3}{2} A'$.  
\label{area}}
\end{figure}

We may parameterize the high field effects of vortex attachment in 
an effective spin Hamiltonian based on the above numerical results 
and our analysis in Sec. IIB.  
Note that the exchange interaction changes sign 
in a roughly periodic fashion as each electron 
captures an additional vortex.  The vortex may be interpreted, 
by equating its Berry's \cite{Berry} phase to an 
Aharonov-Bohm phase, as additional flux.\cite{Jain}  
The confinement, determined by $\omega_0$, fixes the area 
defined by the electronic wave function, $A'$, depicted schematically
by the patterned region on the $N=2$ side of Fig.~\ref{area}.    
From Fig.~\ref{two_R10} we note that we may count the 
number of vortices attached to each electron using the 
flux through the double dot system, $B A'/\Phi_0$.  
The parameter $A'$ is fixed by requiring that  
the net Berry's phase swept out by one quasiparticle (the electron
plus the attached vortices) circling the
other quasiparticle as it encloses the double dot 
system, $2\pi[B A'/\Phi_0-(N-1)p]$,
must vanish for $p=2$.\cite{Jain2}
The net flux includes the effective flux due to the Berry's phase
associated with attaching $p$ vortices to each electron.
The data in Fig.~\ref{two_R10}, 
for example, show that at $B=5.4$ T the flux through the double dot
system exactly cancels the effective flux due to the attached 
vortices (two for each electron).  
By fixing $A'$ in this way, $\pi B A'/\Phi_0$ increases by integer 
multiples of $\pi$ as each electron captures an additional vortex.  
We may then write a parameterized spin Hamiltonian (up to second order in $t^2/U$):       
\begin{eqnarray}
H_{\text{eff}}^{(2)}=2 \tilde{J}_{12} \textbf{S}_1\cdot \textbf{S}_2,
\end{eqnarray}
where $\tilde{J}_{12} \sim \frac{2t^2}{U}\text{cos}\left(\frac{\pi
  BA'}{\Phi_0}\right)$.  We determine $A'$ from our numerical data and set $t=\vert
t_{12} \vert$.  
We have, for simplicity, excluded the 
Zeeman term.  From
Fig.~\ref{two_R10}, for example, we find 
the center of the $p=2$ region to be $\Phi_0/A'\approx 5.4$T which 
gives $A'\sim 800$ $\text{nm}^2$.  The parameter $A'$ suggests 
confinement of an appreciable part of the single electron 
density to within a radius of $\sim 10$ nm.  When we insert 
the magnetic field dependence \cite{Loss2} of $\vert t \vert$ and $U$ into 
$H_{\text{eff}}^{(2)}$ we obtain qualitative agreement with 
our numerical estimates of $J_{ij}$ at {\em all} magnetic 
fields.  But, without the cosine term, $\frac{t^2}{U}$ remains positive for all $B$.

\subsection{Three Quantum Dots}

We now study the $N=3$ system where the quantum dots lie 
at the vertices of an equilateral triangle with side lengths $R=40$nm. 
\cite{chiral}  
We know from the previous section that a large inter-dot separation 
will prevent unwanted excited states of the quantum 
dot from approaching the spin states defining our 
qubit.  We further expect the analysis of Sec. IIA 
to hold only for low magnetic fields while, at large fields, 
electrons capture vortices and initiate spin transitions.  
As a consequence, an external magnetic field has three 
noticeable effects.  
1).  At low fields the length scale, $a$,  is set by confinement and the 
flux enclosed by the triangular loop will dominate the magnetic field 
dependence of the states in Eq.~(\ref{Nthree}).  2).  At higher fields the length scale shrinks with increasing
magnetic field.  The inter-dot tunneling matrix elements will be
suppressed as the electrons become more localized on each dot. 
3).  
The electrons will simultaneously capture vortices to screen the increased 
Coulomb interaction.  The latter effect, as for the $N=2$ system, 
should, in the appropriate parameter regime, lead to oscillations 
in the total spin of the ground state as a function of magnetic field. 

\begin{figure}
\includegraphics[width=2.3in]{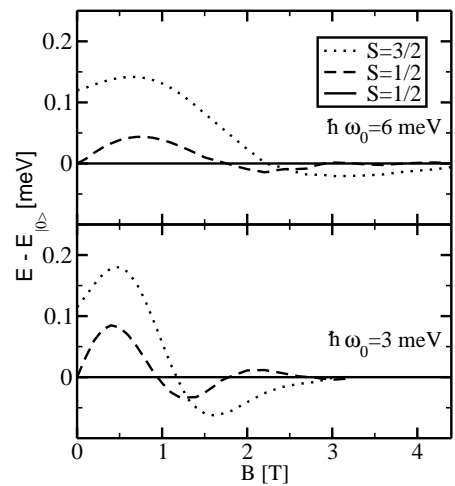}
\caption{  
Energy of the three lowest energy states versus perpendicular magnetic
field obtained from exact diagonalization of Eq.~(\ref{FullH}) in 
the $S_z=1/2$ sector for $N=3$.  The centers of the lateral, parabolic
quantum dots lie at the vertices of an equilateral triangle with
$40$ nm side lengths.  The dotted line has total spin $S=3/2$ while
the dashed ($\vert \lambda=1\rangle_3$) and solid lines ($\vert \lambda=0\rangle_3$) 
have $S=1/2$.  The energy of the $S=1/2$ state corresponding 
to $\vert 0\rangle_3$ is set to
zero.  The top panel has a parabolic confinement parameter
$\hbar\omega_0=6$ meV while the bottom panel has $\hbar\omega_0=3$
meV.  The change in confinement changes the effective area of the
system and, as a consequence, the position of degeneracy 
points between $B=1$ T and $3$ T.
\label{three}}
\end{figure}

Fig.~\ref{three} shows the energy of the lowest states obtained
from exact diagonalization of $H$ in the $S_z=1/2$ sector as a
function of magnetic field.  The confinement parameter is 
taken to be $\hbar\omega_0=6$ meV for the top panel and $3$ meV for 
the bottom panel.  The energy of the state 
$\vert\lambda=0\rangle_3$ is set to zero.  At $B=0$ the two lowest energy states have 
total spin $S=1/2$ and are degenerate, as expected from the 
reflection symmetry of the triangular confining potential.  The 
next highest state has $S=3/2$ which corresponds to $6 \vert t_{ij}
\vert^2/U\approx 0.13$ meV.  
Above this state we find (not shown) the higher excited states 
to lie above $1$ meV.     

We focus first on the low magnetic field data.  
As we increase the magnetic field the
magnetic vector potential breaks the 
symmetry of the confining potential 
leading to a splitting between the two
lowest states.   The splitting is linear in $B$, for small $B$, 
as in Eq.~(\ref{split}).  
We expect such a simple behavior because the two body 
terms in Eq.~(\ref{eff}) have weak magnetic field dependence at low 
magnetic fields, through $\vert t_{ij} \vert^2/U$.
Alternatively, the chiral 
term annihilates $S=3/2$ states.  
Only two body terms in Eq.~(\ref{eff}) affect 
the $S=3/2$ state.  We therefore expect that the energy of
the $S=3/2$ state, $E_{S=3/2}$, decreases very slowly with increasing magnetic 
field at low fields (while  
$E_{S=3/2}-E_{\vert0\rangle_{3}}$
should increase linearly).  
Here the contribution from the chiral term is
sizable and is, for the parameters studied here, larger than the 
Zeeman splitting of a single spin in GaAs, $\approx 0.025 B[T]$ meV.

We now turn to the high field effects in Fig.~\ref{three}.
For highly localized, non-interacting particles 
we expect the flux in $\mathcal{B}_{123}$ to be 
$B A/\Phi_0$, where $A$ is simply the 
the area of the triangle defined by the centers of the  
three quantum dots.  
However, in our system, the parabolic confinement will not perfectly localize the 
interacting electrons.  The area swept out by a closed loop around the 
bulk of the wave function will enclose an area larger than the 
triangle.  Fig.~\ref{area} shows a schematic representation of the 
new, larger area encompassed by the $N=3$ system.  The additional area
due to the expansion of the wave function is $\approx 3A'/2$, where
$A'$ is the area enclosed by an equivalent $N=2$ system.  We may use our 
analysis from the previous section to determine the area added to the triangle.   
The net Berry's phase associated with virtual tunneling processes 
of quasiparticles around the $N=3$ system will be:
\begin{eqnarray}
2\pi\left(\frac{BA}{\Phi_0}+\frac{3BA'}{2\Phi_0}-(N-1)p\right).
\label{phase}
\end{eqnarray}
The additional flux will appear in the 
flux dependent factors in Eq.~(\ref{eff}).   We rewrite the three spin
Hamiltonian in the case of equal tunneling, $t$
(excluding the Zeeman, extended Hubbard, and fourth order terms): 
\begin{eqnarray}
&&H_{\text{eff}}^{(3)}=
\tilde{J}\sum_{i,j}\textbf{S}_i\cdot \textbf{S}_j
\nonumber
\\
&+&\frac{24t^3}{U^2}\text{sin}\left(\frac{2\pi \Phi_{(3)}}{\Phi_0}\right) 
\textbf{S}_1\cdot \textbf{S}_2\times\textbf{S}_3,
\end{eqnarray}
where $\tilde{J}\sim\frac{2t^2}{U}\text{cos}
\left(\frac{\pi\Phi_{(2)} }{\Phi_0}\right)$.  We 
define $\Phi_{(2)}=BA'$ and $\Phi_{(3)}\approx B(A+3A'/2)$. 
Note that integer $p$, in Eq.~(\ref{phase}), does not contribute to the chiral term.
Therefore, vortex attachment does not directly affect the flux in the chiral term.
Furthermore, the cosine in the first term parameterizes large
magnetic field behavior while the sine in the second, chiral term, 
was derived using perturbation theory.  
$H_{\text{eff}}^{(3)}$ allows us to
predict the degeneracy point of the three spin term using the
degeneracy point of the two spin term with the following energies: 
\begin{eqnarray}
E_{S=3/2}-E_{\vert 0\rangle_3}&=&\frac{6t^2}{U}\text{cos}\left(\frac{\pi \Phi_{(2)}}{\Phi_0}\right)
\nonumber
\\
&+&\frac{6\sqrt{3}t^3}{U^2}\text{sin}\left(\frac{2\pi 
\Phi_{(3)}}{\Phi_0}\right)
\label{split1}
\\
E_{\vert 1\rangle_3}-E_{\vert 0\rangle_3}
&=&\frac{12\sqrt{3}t^3}{U^2}\text{sin}\left(\frac{2\pi \Phi_{(3)}}{\Phi_0}\right).
\label{split2}
\end{eqnarray}
From the dotted line 
in the top panel of Fig.~\ref{three} and Eq.~(\ref{split1}) we 
find $\Phi_0/4A '\approx 2.2T$.  Using Eq.~(\ref{split2}), 
we predict the dashed line to
cross the $x$-axis near $B\approx 1.5T$, 
where we take $A$ to be the area
of an equilateral triangle with $40$ nm side lengths.  
A similar analysis yields good agreement for the 
bottom panel of Fig.~\ref{three}.         

The slope of the energy splitting between the two lowest states in 
Fig.~\ref{three} allows us to estimate $t/U$ for this system using 
Eq.~(\ref{split1}).  
$t/U$ is largest and only weakly magnetic field dependent at low $B$.
Taking $A+\frac{3}{2}A'$ from above
and $v=0$, we obtain $t/U\simeq 0.09$ for the top panel and 
$0.19$ for the bottom panel which shows that our 
expansion in $t/U$ is consistent.  For $N=3$, 
only odd powers of $t_{ij}$ allow linear magnetic 
field dependence in the splitting, showing
that, excluding double occupancy, the magnetic field dependence captured by the chiral term 
is accurate up to $\vartheta\left(\frac{t^5}{U^4}\right)$. 

\section{Conclusion} 

We show in this work that the Heisenberg model description of the
qubit coupling in the quantum dot spin quantum computer architecture
applies only in a limited regime of the parameter space.  In the GaAs
quantum dot exchange gate architecture, the Heisenberg spin
Hamiltonian description applies only in the intermediate regime
$R/a\gtrsim 1$ and $\omega_c/\omega_0\lesssim 1$.  Using the exact
diagonalization technique one can map out the precise low lying
Hilbert space, and consequently, use this information in the design of
the quantum computer architecture.  We have also discussed an
interesting and non-trivial level-crossing periodicity in the
singlet-triplet energetics.  Precise knowledge of the 
associated, low-energy Hilbert space could, in principle, be used to
protect quantum information encoded in the electron spin.  

Generalizing our exact diagonalization technique to spin cluster
qubits formed by a two-dimensional array of electron spins localized
in tunnel-coupled quantum dots, we show that the chiral term
associated with the quantum phase picked up by an electron enclosing
the magnetic flux through closed loops must be included in 
the spin Hamiltonian.  The existence of the chiral term in the looped
spin cluster qubits modifies the Heisenberg interaction and is in some
sense a decoherence mechanism for the simple exchange gate
architecture (since the two-qubit SWAP operation is no longer 
determined by just the Heisenberg exchange Hamiltonian).  We show in this
paper how precise knowledge of the cluster geometry, combined with exact
diagonalization, provides us with the multi-spin Hamiltonian which
would be required for quantum computation with two or three
dimensional spin cluster qubits.  Strictly one dimensional spin
cluster qubits, which do not have any topological looping, have a
small (but non-zero) chiral contribution and are therefore described,
for the most part, by the Heisenberg Hamiltonian in the appropriate
subspace of magnetic field, confinement, and dot geometry parameters.

We emphasize that in this article we have considered a relatively
simple model, defined by Eqs.~(\ref{FullH}) and (\ref{pot}), for
determining the applicability of the Heisenberg interaction in
describing the exchange gate operation.  Differences in the
confinement potential may change some of the quantitative aspects of
our results but as long as the confinement consists of smooth
potential wells, there should be 
qualitative agreement.  The key issues we have
addressed in this work is the regime of validity of the Heisenberg
exchange gate operation in coupled semiconductor quantum dot quantum
computer architectures as appropriate, for example, in GaAs-based 
quantum dot systems.  In practice, we have obtained the conditions and
constraints necessary for a coupled qubit system to behave as a
coherent molecule as opposed to two decoupled atoms.  Adiabatic tuning
between these two regimes enables the swap idea underlying the
exchange gate.  Of course, the issue of He atom-to-molecule transition
in coupled quantum dot systems as well as our discussion of the
exchange oscillations in the coupled dot system as a function of the
applied magnetic field have implications beyond quantum computation.
For example, a direct experimental observation of the exchange
oscillations is of interest in quantum dot physics.

Finally we mention that there are many other factors beyond the scope
of our work ({\em i.e.} beyond the model defined in Eqs.~(\ref{FullH}) 
and (\ref{pot})) which affect the operation of the exchange gate.  We
cite three such example of recent interest which have been considered
in the literature: Inhomogeneous magnetic field effects \cite{Hu2},
spin-orbit coupling \cite{Kavokin,Bonesteel}, and multi-valley quantum
interference \cite{Koiller}.

We would like to thank J.K Jain and K. Park for many helpful
discussions.  This work is supported by ARO-ARDA and NSA-LPS.


\end{document}